\documentstyle[12pt]{article}
\title{Charge inversion in DNA--amphiphile complexes: Possible application to gene 
therapy}
\author{ Paulo S. Kuhn,  Yan Levin\footnote{Corresponding author: 
levin@if.ufrgs.br}, and Marcia C. Barbosa \\
Instituto de F\'{\i}sica, Universidade Federal
do Rio Grande do Sul\\ Caixa Postal 15051, CEP 91501-970, Porto
Alegre, 
RS, Brazil}

\begin{document}
\maketitle
\begin{abstract}

We study a complex formation between the  DNA and cationic amphiphilic molecules.  
As the  amphiphile is added to the solution containing DNA, 
a cooperative binding of surfactants to the DNA molecules is found. 
This binding transition occurs at 
specific density of amphiphile, which is strongly dependent on the concentration
of the  salt and on the hydrophobicity of the surfactant
molecules.   We find that for amphiphiles which
are sufficiently hydrophobic, a charge neutralization, or even charge inversion 
of the complex is possible.
This is of particular  importance in applications to gene therapy, for
which the functional delivery of specific base sequence into living cells 
remains an outstanding problem. 
The charge inversion could, in principle, allow the DNA-surfactant 
complexes to approach  negatively charged cell membranes permitting 
the transfection to take place.

\end{abstract}

\bigskip

PACS.05.70.Ce - Thermodynamic functions and equations of state

PACS.61.20.Qg - Structure of associated liquids: electrolytes, molten salts, etc.

PACS.61.25.Hq - Macromolecular and polymer solutions; polymer melts; swelling

\newpage

\section{Introduction}

In the last few years gene therapy has received significant
 attention both from the scientific 
community and from the general public.  The development of new techniques for
 transferring genes into living cells allowed for the potential 
treatment of several diseases of genetic origin \cite{Fried}-\cite{Hope}. 
The central problem of gene therapy is the development of safe and efficient 
gene delivery system. Since both the $DNA$ and the cell membranes are 
negatively charged, the naked polynucleotides are electrostatically prevented from
entering inside cells. Furthermore, the unprotected $DNA$ is rapidly degraded by
nucleases present in plasma \cite{Hope}. 

 Although, much effort has concentrated on viral 
 transfection, non-viral methods have received increased attention.  This
is mostly due to the possible complications which can arise from recombinant viral 
structures, and the consequent risk of cancer. In the non-viral category, the
$DNA-liposome$ complexes have shown the  most promise. Cationic liposomes
 can associate with the $DNA$ segments, neutralizing or even 
inverting the electric charge
 of nucleotides, thus significantly  increasing the efficiency of  gene adsorption
and transfection by cells.

In this paper we present a model of $DNA-amphiphile$ solutions. 
We find that in equilibrium,
solution consists of complexes composed of DNA and associated 
counterions and amphiphiles. As more amphiphiles are added to solution, 
a cooperative binding transition is found. At the transition point 
a large fraction
of the $DNA's$ charge is neutralized by the condensed surfactants.  If the density of
surfactant is increased beyond this point, a charge inversion of the DNA becomes
possible.  The necessary density of amphiphile needed to reach the
charge inversion is strongly dependent on the characteristic hydrophobicity 
of surfactant molecules.  In particular, we find that for sufficiently hydrophobic
amphiphiles, such as for example some cationic lipids, the charge inversion can
happen at extremely low densities.

\section{The model}

Our system  consists of an aqueous solution of  $DNA$ segments, 
cationic surfactants, and
monovalent salt. Water is modeled as a uniform medium of dielectric constant $D$. 
In an aqueous solution, the phosphate groups of the $DNA$ 
molecules become ionized resulting in a net negative charge.  
The salt is completely ionized, forming 
 an equal number of cations  
and anions. 
Similarly the surfactant molecules are assumed to be fully dissociated
producing negative anions and polymeric chains with cationic 
head groups.

Following the usual nomenclature, we shall call the 
ionized $DNA$ molecules the ``polyions'', the positively
 charged ions the ``counterions'', 
and the negatively charged anions the ``coions''. To simplify the calculations, 
all the counterions and coins will be
treated as identical, independent of the molecules from which they were derived.
The  $DNA$ strands will be modeled 
as long
rigid cylinders of length $L$ and
diameter $a_p$, with the  
charge $-Zq$ distributed  
 uniformly, with separation $b \equiv
L/Z$,  along the major axis. 
The cations and anions will be depicted as hard spheres of diameter $a_c$  and
charge $\pm q$. For simplicity we shall also suppose that  each one of the $s$
surfactant monomers is a rigid sphere of diameter $a_c$ with the ``head'' monomer
carrying the charge $+q$. The interaction between the  
hydrophobic tails is short ranged and characterized by the hydrophobicity
parameter $\chi$ 
(see Fig. $1$).  The density of $DNA$
segments is $\rho_{p}=N_p/V$, the 
density of monovalent salt is $\rho_m=N_m/V$, and the density of amphiphile
is $\rho_s=N_s/V$, where $N_i$ 
is the number of 
molecules of specie $i$ and $V$ is the volume of the system.

The strong electrostatic attraction between the polyions, counterions,
and 
amphiphiles, leads to formation of complexes consisting 
of {\it one} polyion, $n_c$ counterions, and $n_s$ amphiphilic 
molecules. We shall assume that to each phosphate
group of the $DNA$ molecule can be associated at most {\it one} counterion {\bf or}
$l \le l_{max}$ surfactants.  This assumption seems to be
quite reasonable in view of the fact that 
the electrostatic repulsion between the counterions will prevent more than one
counterion from condensing onto a given monomer.  On the other hand, the 
gain in hydrophobic energy resulting from the close packing of the surfactant
molecules might be able to overcome the repulsive electrostatic  
interaction between the surfactant head groups, favoring condensation of more than
one surfactant on a given monomer (see Fig. $2$). The  $l$ amphiphilic molecules form a ``ring'' of 
radius $a$ around the central negative monomer of the $DNA$ (see Fig. $3$). 
If we assume that 
most of the hydrocarbon chain of the associated surfactants 
is hidden inside the $DNA$ molecule, the maximum number of surfactants in a ring 
can be estimated from
the excluded volume considerations, $l_{max}=2 \pi a /a_c$, 
where $a \equiv (a_p + a_c)/2$ is the radius of the exclusion 
cylinder around a polyion.

At equilibrium, each site (monomer) of a polyion can be free  or  
have  one counterion {\it or} a ring of $l=1,...,l_{max}$ 
surfactants associated to it.
We define the surface coverage of counterions as
$p_c=n_c/Z$, and the surface coverage of surfactant rings 
as $p_l=n_l/Z$, where $n_c$ is the number of condensed counterions and
$n_l$ is the number of rings containing  $l$ surfactants. 
Each polyion has a distribution of 
rings containing from one to $l_{max}$ surfactants.
We shall neglect the polydispersity in the size 
of the complexes, assuming that all the  complexes have $n_c$
counterions and $n_s$  amphiphilic 
molecules --- in rings of $\{p_l\}$ --- with 
\begin{equation}
\label{eq00}
n_s=\sum_{l=1}^{l_{max}}Z l p_l \; .
\end{equation}
The total charge of each polyion  
is, therefore, renormalized from $-Zq$ to $-Z_{eff}q$, with 
$Z_{eff}\equiv Z-n_c-n_s$ \cite{Alex}-\cite{LevBarb1}. 
>From overall charge neutrality, the density
of free  cations is $\rho_+=\rho_m+(Z-n_c)\rho_p$,
the density of free anions is $\rho_-=\rho_m+\rho_s$, and
the density of free surfactants is $\rho_s^f=\rho_s-n_s\rho_p$.
 We shall restrict our attention to 
the limit of low surfactant densities, 
so as to prevent micellar formation in the bulk.

The aim of the theory is to determine the characteristic 
values of $n_c$, $n_s$, and 
the surface coverage by rings $\{p_l\}$. 
To accomplished this, the free energy of the $DNA-surfactant$ solution
will be constructed and minimized.

\section{The Helmholtz free energy}
 
 The free energy is composed of three contributions,
\begin{equation}
\label{eq0}
F=F_{complex}+F_{electrostatic}+F_{mixing} \; .
\end{equation}
The first term is the free energy needed to form the  isolated 
complexes.  The second term accounts for the electrostatic interaction
between the counterions, coions, surfactants and complexes. Finally, 
the third term
is the result of entropic mixing of various species.

To calculate the free energy required to construct an isolated complex composed of
one polyion, $n_c$ condensed counterions, and $n_s$ condensed surfactants, 
we employ the following simplified model.  Each 
monomer of a polyion can be free or occupied 
by a 
counterion,  {\it or} by  $1\leq l\leq l_{max}$ amphiphiles (see Fig.
$2$). 
Therefore,  to each monomer $i$  
we  associate occupation variables $\sigma_c(i)$ and $\{\sigma_l(i)\}$,
which are nonzero if that particular monomer is occupied by a condensed 
counterion or a 
ring with $l$ surfactants, respectively. 
The free energy of $N_p$ isolated complexes can then be written as
\begin{equation}
\label{eq1}
\beta F_{\rm complex} = - N_p\ln \sum_{\nu}^{} e^{-\beta E_{\nu} } \, ,
\end{equation}
where the sum is over all possible configurations of counterions and
surfactants along a complex. For
a particular 
configuration $\nu$, the energy can be expressed as the  sum of three terms,
$E_{\nu}=E_1+E_2+E_3$. The first one
is the 
electrostatic contribution arising from the Coulombic interactions
between all 
charged sites of a complex,
\begin{equation}
\label{eq2}
E_1 = \frac {q^2} {2} \sum_{i \neq j}^Z \frac {[ - 1 + \sigma_c(i) +
\sum_l^{l_{max}} l \sigma_l(i)] [ - 1 + \sigma_c(j) +
\sum_l^{l_{max}} l \sigma_l(j)]} {D |r(i)-r(j)|} \, ,
\end{equation}
 where we have assumed that the only effect of association is the 
renormalization of the effective charge of each monomer.  
The second term $E_2$,
is due to  
hydrophobic interactions between the surfactant molecules,
\begin{equation}
\label{eq3}
E_2 = \frac{\chi}{2} \sum_{\langle i,j \rangle}^Z\sum_{l,l'=1}^{l_{max}} \frac {(l+l')} {2} \sigma_l(i) \sigma_{l'}(j) \; ,
\end{equation}
where, in order to simulate the short-ranged nature of hydrophobic interactions,  
the first sum is constrained to run over the nearest neighbors.
The 
hydrophobicity parameter $\chi$ is negative, representing the
tendency of the two adjacent surfactant molecules to expel water.
We can estimate its value from the experimental measurement of the energy
necessary to 
remove an amphiphile from a monolayer and place it in the bulk \cite{Isra}.

The third contribution $E_3$, accounts for the internal energy of each ring,
\begin{equation} 
\label{eq4}
E_3 = \sum_{i}^Z \sum_{l=2}^{l_{max}}\sigma_l(i)E_l \; .
\end{equation}
$E_l$ is the interaction energy between $l$ surfactants forming a ring. 
Each ring contains a maximum
of $l_{max}$ sites, which can be occupied by surfactants. 
To each one of these sites we associate an
occupation variable  $\tau(j)$, which is zero if site $j$ is unoccupied by
a surfactant and is one if it is occupied ( see Fig. $3$). The interaction 
energy
of surfactants forming a ring can then  be written as 
\begin{equation} 
\label{eq5}
E_l=\frac{q^2}{2 D}\sum_{i\neq j}^{l_{max}}\frac{\tau(i)\tau(j)}
{2 a \sin(\pi|i-j|/l_{max})}+ \frac {\chi} {2} \sum_{\langle i,j \rangle}^{l_{max}} \tau(i) \tau(j) \; .
\end{equation}
The first term of Eq.(\ref{eq5}) is due to  electrostatic repulsion between
the surfactant head groups, while the second is the result of attraction between
the adjacent hydrocarbon tails.

The exact solution  of even this simpler sub-problem 
(i.e. evaluation of the sum in Eq.(\ref{eq1})) is very difficult 
due to the long ranged electrostatic interactions. We shall, therefore,
resort to mean-field theory, which  works  particularly well
for long-ranged potentials.   Evaluating 
the upper bound for the free energy, given by the
Gibbs-Bogoliubov inequality, and neglecting the end effects
we obtain,
\begin{eqnarray}
\label{eq6}
\beta F_{\rm complex} & = &\beta N_p[ f_{el}+ f_{hyd}+ f_{ring}+ f_{mix}] \; .
\end{eqnarray}
The first term, 
\begin{eqnarray}
\label{eq7}
\beta f_{el}&=& \xi S 
[-1+ p_c + \sum_{l=1}^{l_{max}}l p_l]^2-\xi S N_p \; ,
\end{eqnarray}
is  the electrostatic interaction between the 
sites along one rod and is related to $E_1$. $S$ is expressed in terms of 
the digamma function \cite{GR},
\begin{eqnarray}
\label{eq8}
S&=&Z[\Psi(Z)-\Psi(1)]-Z+1 \; ,
\end{eqnarray}
and  $\xi \equiv \beta q^2/Db$ is 
the Manning parameter \cite{Man1,Joan}. 
The second term in Eq.~(\ref{eq6}), 
\begin{eqnarray}
\label{eq9}
\beta f_{hyd}&=&\beta \chi (Z-1) \sum_{n,m}^{l_{max}} \frac{(n+m)}{2}p_m p_n \; ,
\end{eqnarray}
is the hydrophobic attraction between the rings inside
a complex. The third term is the free energy due to the electrostatic and
hydrophobic interactions between the surfactants forming a ring,
\begin{eqnarray}
\label{eq10}
\beta f_{ring} & = & \frac{2\ln l_{max} + \nu_0}{4\pi T^*} \sum_{l=2}^{l_{max}}Zp_l l^2 
 + \frac {\beta \chi} {l_{max}} \sum_{l=2}^{l_{max}} Z p_l l^2 +  \\
 && \sum_{l=1}^{l_{max}}
Z p_l l \ln \left( \frac {l} {l_{max}} \right) + \sum_{l=1}^{l_{max}} Zp_l l_{max} \left( 1-\frac{l}{l_{max}} \right) 
\ln \left( 1-\frac{l}{l_{max}} \right) \nonumber  
\end{eqnarray}
where $\nu_0\approx 0.25126591$, and the reduced temperature is $T^*=k_BTDa/q^2$.
Finally, the free energy of mixing for rings and counterions of a complex is,
\begin{eqnarray}
\label{eq11}
\beta f_{mix}&=&Z  (1-p_c-\sum_{l}^{l_{max}}p_l)+
\ln (1-p_c-\sum_{l=1}^{l_{max}}p_l) +Z   p_c \ln p_c + \nonumber  \\
&& Z\sum_{l=1}^{l_{max}}p_l \ln p_l- Z p \ln l_{max} + \\  
&& Z p l_{max} \left(1 - \frac {1} {l_{max}} \right) 
\ln \left(1 - \frac {1} {l_{max}} \right)   \nonumber  \; , 
\end{eqnarray}
where to be consistent with the expression (\ref{eq10}), we have included
a contribution to the free energy  arising 
from the azimuthal motion of condensed counterions around the polyion, {\it i.e.} 
the last two 
terms of Eqn.(\ref{eq11}). 

Once a cluster, constructed in isolation, is introduced into  solution,
it gains an additional solvation energy due to its interaction with other clusters,
free counterions, free coions, and free surfactants.  The 
electrostatic repulsion between the complexes is screened by the ionic atmosphere, 
producing  an 
effective short ranged potential of DLVO form \cite{Der}-\cite{FisLevLi}.
The electrostatic free energy due to interactions
between various clusters can be estimated from the second virial coefficient,
\begin{eqnarray}
\label{eq12}
\beta F^{cc} = (Z - n_c - n_s)^2 \frac {2 \pi N_p^2 a^3 e^{-2 \kappa a}} {V T^* (\kappa a)^4 K_1^2(\kappa a)} \, ,
\end{eqnarray}
where $ (\kappa a)^2 \equiv 4 \pi \rho^*_1/ T^*$ and
$\rho_1^* \equiv a^3[\rho_p(Z-n_s-n_c)+2\rho_m+2\rho_s]$ is 
the reduced density of free ions.   
The free energy due to interaction between the complexes and  
 free ions and  surfactants can be obtained following the general
methodology of the Debye-H{\"u}ckel-Bjerrum theory\cite{FL1,FL2}, \cite{DH1}-\cite{Surfoplex},
\begin{eqnarray}
\label{eq14}
\beta F^{ci} & = & N_p (Z - n_c - n_s)^2 \frac {(a/L)} {T^* (\kappa a)^2} \times \nonumber \\
& & \times \left[ -2 \ln (\kappa a K_1(\kappa a)) + I(\kappa a) 
- \frac {(\kappa a)^2} {2} \right] \, ,
\end{eqnarray}
with
\begin{equation}
\label{eq15}
I(\kappa a) = \int^{\kappa a}_0 \frac {x K_0^2(x)} {K_1^2(x)} dx \, ,
\end{equation}
where $K_n$ is the modified Bessel function of order $n$. The contribution 
to the total free energy arising from the interactions
between the free ions and surfactants is given by the usual Debye-H\"uckel expression
\cite{DH1},\cite{DH2}
\begin{equation}
\label{eq13}
\beta F^{ii} = - \frac {V} {4 \pi a_c^3} \left[ \ln(1 + \kappa a_c) -
\kappa a _c + \frac {(\kappa a_c)^2} {2} 
\right] \, .
\end{equation}
This term is very small and is included only for completeness.

The last contribution to the total free energy Eq.~(\ref{eq0}), 
results from the entropic mixing of 
the counterions, 
coions, surfactant and complexes, 
\begin{equation}
\label{eq16}
F_{mixing}=F_{m+}+F_{m-}+F_{s}+F_{c} \, .
\end{equation}
The free energy of mixing is obtained following the general ideas introduced
by Flory \cite{Flo},
\begin{eqnarray}
\label{eq17}
\beta F_{m+} & = & N_{m+} \ln \phi_{m+} - N_{m+} \, , \nonumber \\
\beta F_{m-} & = & N_{m-} \ln \phi_{m-} - N_{m-} \, , \nonumber \\
\beta F_{s} & = & N_{s} \ln (\phi_{s}/n_s) - N_{s} \, , \nonumber \\
\beta F_{c} & = & N_{p} \ln \left( \frac {(Z + n_c + n_s) 
\phi_{c}} {Z + n_c + n_s s} \right) - N_p \, .
\end{eqnarray}
In the above expression $m+$ denotes free counterions, $m-$
free coions, $s$ free surfactant molecules, and $c$ 
complexes. The
\begin{eqnarray}
\label{eq18}
& & \phi_{m+} = \frac {\pi \rho^*_+} {6} \left( \frac {a_c} {a} \right)^3 \, , \nonumber \\
& & \phi_{m-} = \frac {\pi \rho^*_-} {6} \left( \frac {a_c} {a} \right)^3 \, , \nonumber \\
& & \phi_{s} = \frac {s \pi \rho_s^{f*}} {6} \left( \frac {a_c} {a} \right)^3 \, , \nonumber \\
& & \phi_{c} = \pi \rho_{p}^* \left[ \frac {1} {4 (a/L)} \left( \frac {a_p} {a} \right)^2 + \frac {1} {6} (n_c + n_s s) \left( \frac {a_c} {a} \right)^3 \right] \; 
\end{eqnarray}
are the volume fractions occupied by  the free  counterions, coions, surfactants,  and
complexes, respectively.
\bigskip

\section{Results and Conclusions}

The  equilibrium configuration of the polyelectrolyte-surfactant solution 
is determined by the 
requirement that the Helmholtz free energy  be minimum.
Since $F$ is the function of $n_s,n_c$, and the surface coverage by
rings $\{p_l\}$, minimization of $F$ implies 
that
\begin{eqnarray}
\label{eq19}
\delta F=\frac{\partial F}{\partial n_s}\delta n_s+ 
\frac{\partial F}{\partial n_c}\delta n_c+
\sum_{l=1}^{l_{max}}\frac{\partial F}{\partial p_l}\delta p_l=0\;\; .
\end{eqnarray}
Using the constraint Eq.~(\ref{eq00}), Eq.~(\ref{eq19}) can
be separated into $l_{max}+1$ equations,
\begin{eqnarray}
\label{eq20}
 \frac{\partial F}{\partial n_c}=0
\end{eqnarray}
and
\begin{eqnarray}
\label{eq21}
 \frac{\partial F}{\partial n_s}Zl+
\frac{\partial F}{\partial p_l}=0 \; ; \; \;  l=1...l_{max}.
\end{eqnarray}
The system of equations (\ref{eq20}) and (\ref{eq21}) can, in principle,
be solved numerically.  However, for reasonable values of $l_{max}$ this
requires a significant numerical effort.  Instead of pursuing this
brute force method, we note that to a reasonable accuracy, the surface coverage by rings, $\{p_l\}$, can be 
approximated by an exponential distribution\cite{Lev2},   
\begin{eqnarray}
\label{eq22}
p_l = \frac {n_s e^{\alpha l}} {Z \sum_{l=1}^{l_{max}} l e^{\alpha l}} \; .
\end{eqnarray}
We have checked that this is, indeed, a good approximation by numerically
solving Eq.~(\ref{eq21}) for an isolated complex.
Using ansatz (\ref{eq22}), the total free energy 
becomes a function of $n_c$, $n_s$, and $\alpha$. For a fixed volume and number
of particles the equilibrium corresponds to the minimum of Helmholtz
free energy,
\begin{eqnarray}
\label{eq23}
\frac{\partial F}{\partial n_c}=0 \, , \\  
\frac{\partial F}{\partial n_s}=0 \, , \\  
\frac{\partial F}{\partial \alpha}=0 \; .
\end{eqnarray}
These are three coupled algebraic equations, which can be easily
solved numerically to yield the
characteristic number of condensed counterions, surfactants, as well as
the shape of the distribution of ring sizes $(\alpha)$.  
In Fig. 4 we present a numerical solution of these
equations. As a specific example we consider a cationic surfactant with
an alkyl chain of $s=12$ groups. In this case
the hydrophobicity parameter can be estimated \cite{Surf} to be in the range of
$\chi \approx -3,5 k_B T$. To explore the dependence of condensation
on the hydrophobicity of surfactant, we shall vary  this value within reason.
The density of monovalent salt 
and the $DNA$ is taken  to be $18\;mM$ and $2 \times 10^{-3}mM$, respectively.

The resulting  binding isotherms are illustrated  in 
Fig. $4$. The fraction of associated amphiphilic molecules
$\beta_s=n_s/Z$, is plotted against the density of surfactant for
 a fixed amount of monovalent salt, $\rho_m$.
For small concentrations of cationic surfactant, few
amphiphilic molecules associate with the $DNA$ segments.  At the certain
critical concentration, however, the system forms {\it surfoplexes} \cite{Surf}
\cite{Surfoplex} ---  complexes in which the charge of the $DNA$ is
almost completely neutralized by the associated amphiphiles. 
If the density is increased further, on average, more than
one surfactant molecule will associate to each phosphate
group, leading to charge inversion of the surfoplexes. 
 For highly hydrophobic surfactants 
the charge inversion can happen very close to the cooperative binding transition.
We note that our theory predicts the binding transition to be discontinuous,
this, most likely, is an artifact of the mean-field approximation \cite{Surfoplex}.

We have presented a simple theory of $DNA-surfactant$ solutions.
Our results should be of direct interest to researchers working on design 
of improved  gene
delivery systems. In particular we find that addition of cationic
surfactants leads to a strong cooperative binding transition.  This transition
happens far bellow the critical micell concentration.  A further increase
of amphiphile density can result in the charge inversion of the 
$DNA-surfactant$ complexes.  This regime should be particularly useful
in designing gene or oligonucleotide delivery systems.  Until now most of non-viral
gene-delivery systems were in the form of lipoplexes ---  complexes formed
by $DNA$ and cationic liposomes.  To form the liposomes, however, is
required a significant concentration of cationic lipid.  Unfortunately,
at high concentrations
both lipids and surfactants are toxic to organism.
Our model suggests that the  charge inversion can be achieved with  quite small
concentration of cationic amphiphile, {\it if} it is sufficiently hydrophobic. 
This should reduce the risk of unnecessary medical complications. 

\bigskip

{\bf ACKNOWLEDGMENTS}

\bigskip
This work was 
supported in part by  Conselho Nacional de
Desenvolvimento Cient{\'\i}fico e Tecnol{\'o}gico and 
Financiadora de Estudos e Projetos, Brazil.


\newpage

\bigskip
\bigskip
\bigskip
\centerline{\bf FIGURE CAPTION}
\bigskip
\noindent Figure 1. A cylindrical polyion of diameter $a_p$,  length
$L$, and charge  $-Zq$, surrounded 
by  spherical ions of radius $a_c$ and amphiphilic molecules
of $s$ monomers.
 Each monomer of a macroion is free or has  
{\it one} counterion, {\bf or} a ring made of $l$ amphiphilic 
molecules associated with it. 

\bigskip

\noindent Figure 2. Schematic representation of a complex. Empty sites (monomers) 
($-$), sites with associated counterion ($c$), sites with $l$ associated 
amphiphiles ($s_l$).

\bigskip

\noindent Figure 3. Ring composed of $l$ surfactant molecules, $l_{max}=15$.

\bigskip

\noindent Figure 4. Effective binding fraction
of amphiphiles $\beta_{s}\equiv
n_s/Z$, as a function of 
amphiphile concentration $\rho_s$. The concentrations of DNA and of
added salt is $2 \times 10^{-6}M$ 
and $18 mM$, respectively. The length of the DNA segments is $220$ base
pairs.
The solvent is water at room temperature, so that $\xi=4.17$.

\bigskip

\noindent Figure 5. Average size of rings in a complex 
(parameters the same as in Fig. 4).

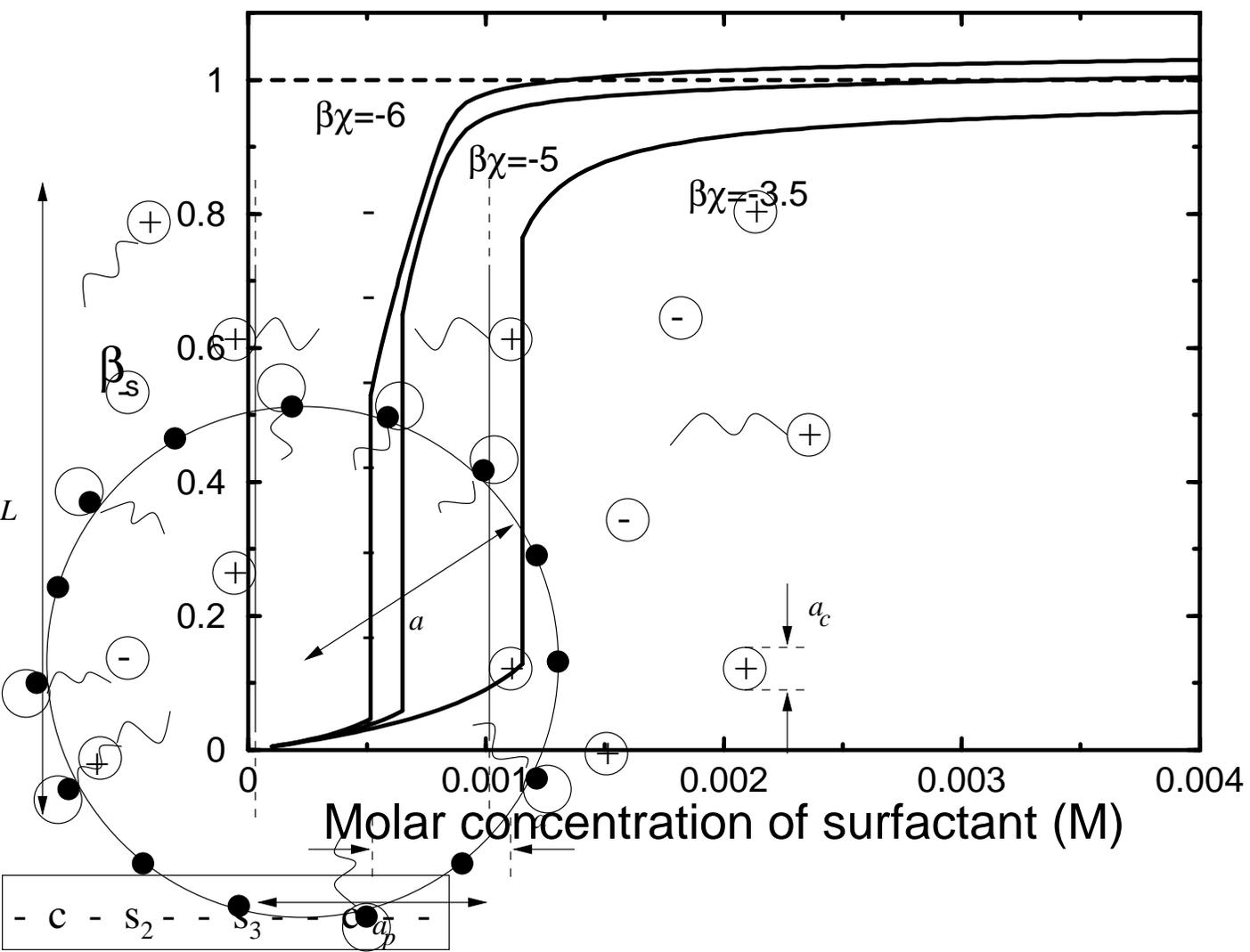
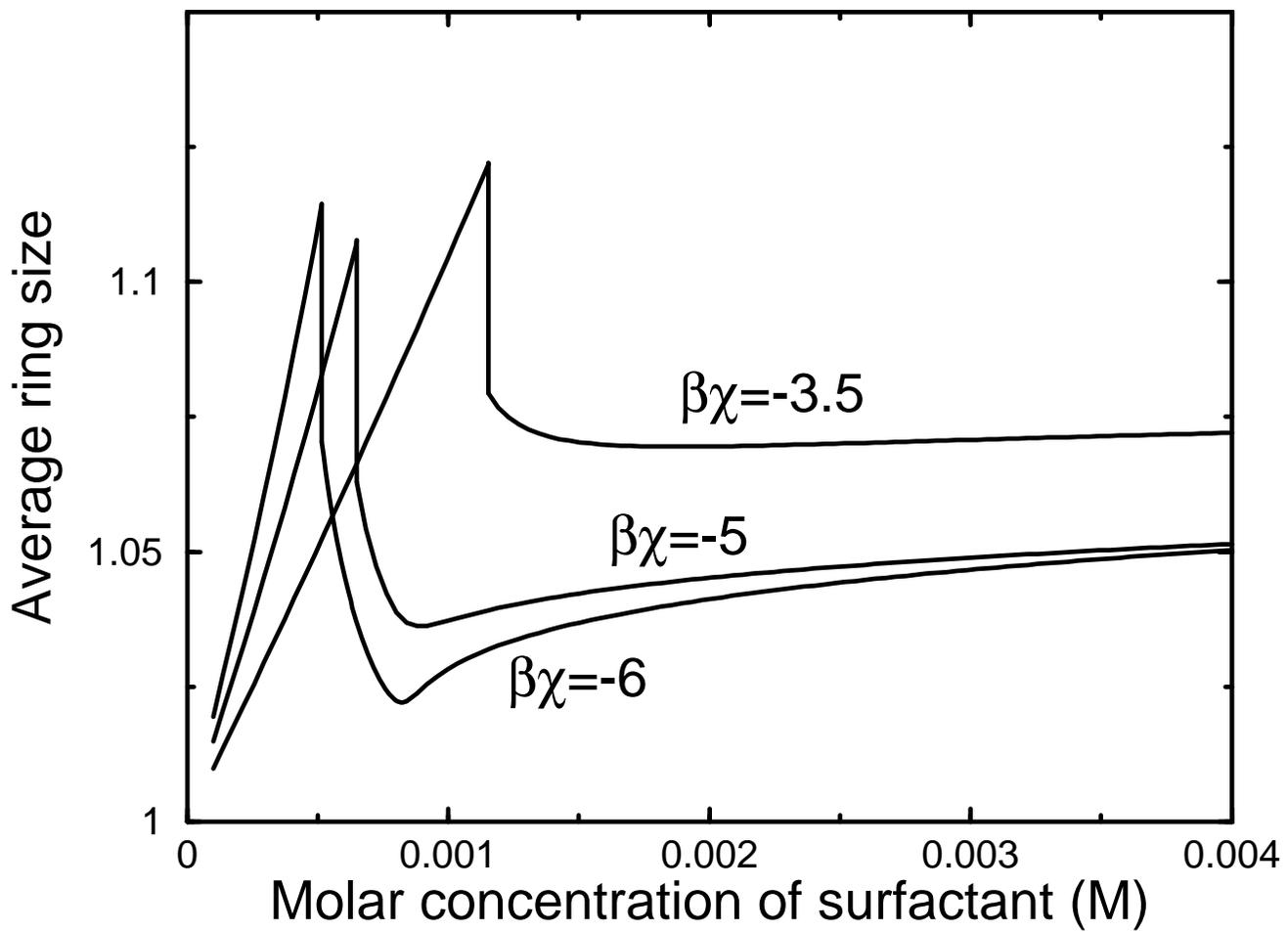
\end{document}